\documentclass[aps, prb,twocolumn,showpacs,floatfix,superscriptaddress,nofootinbib]{revtex4-2}
\usepackage{graphics}
\usepackage{xcolor}
\usepackage{amsmath, amssymb,amsthm}
\usepackage{mathtools}
\usepackage{physics}
\usepackage{bbold}
\usepackage[normalem]{ulem}
\usepackage[colorlinks=true,citecolor=blue,linkcolor=magenta]{hyperref}
\usepackage[caption=false]{subfig}

\newcommand{\diff}[1]{\mathrm{d}{#1}\,}
\renewcommand{\dfrac}[2]{\frac{\mathrm{d}{#1}}{\mathrm{d}{#2}}}

\definecolor{MyBlue}{HTML}{1f77b4}
\definecolor{MyOrange}{HTML}{ff7f0e}
\definecolor{MyGreen}{HTML}{2ca02c}
\definecolor{MyRed}{HTML}{d62728}

\newcommand{\msout}[1]{\text{\sout{\ensuremath{#1}}}}

\newcommand{\mold}[1]{\textcolot{blue}{\msout}}

\begin{document}


\title{Berezinskii approach to disordered spin systems with asymmetric scattering
and application to the quantum boomerang effect}

\author{Jakub Janarek}\email{jakub.janarek@uj.edu.pl}
\affiliation{Instytut Fizyki Teoretycznej,
Uniwersytet Jagiello\'nski,  \L{}ojasiewicza 11, PL-30-348 Krak\'ow, Poland}

\author{Nicolas Cherroret}\email{nicolas.cherroret@lkb.upmc.fr}
\affiliation{Laboratoire Kastler Brossel, Sorbonne Universit\'e, CNRS,
ENS-PSL Research University, Coll\`ege de France, 4 Place Jussieu, 75005
Paris, France}

\author{Dominique Delande}\email{dominique.delande@lkb.upmc.fr}
\affiliation{Laboratoire Kastler Brossel, Sorbonne Universit\'e, CNRS,
ENS-PSL Research University, Coll\`ege de France, 4 Place Jussieu, 75005
Paris, France}


\begin{abstract}
We extend the Berezinskii diagrammatic technique to one-dimensional disordered spin systems, in which time-reversal invariance is broken due to a spin-orbit coupling term inducing left-right asymmetric scattering.
We then use this formalism to theoretically describe the dynamics of the quantum boomerang effect, a recently discovered manifestation of Anderson localization. The theoretical results are confirmed by exact numerical simulations of wave-packet dynamics in a random potential.
\end{abstract}

\maketitle

\section{Introduction}

In the presence of a spatially disordered potential, quantum wave packets may experience, after an transient temporal spreading, a complete freezing of their density distribution due to the proliferation of destructive interference in the multiple scattering process. This phenomenon, which generically occurs in low dimension, is one of the most representative manifestations of Anderson localization \cite{Anderson1958}. As such, it has been primarily exploited in the experimental quest for the localization of cold atoms in random potentials \cite{Billy2008, Roati2008, Semeghini2015, Jendrzejewski2012a}. Recently, however, a variety of alternative signatures of Anderson localization has been identified. Those include the temporal freezing of the coherent backscattering effect in reciprocal space \cite{Cherroret2012, Ghosh2015, Cobus2016} or the universal growth of narrow peak structures in the density profile \cite{Hainaut2017, Hainaut2018} and momentum distribution \cite{Karpiuk2012, Ghosh2014, Loon2014}  of spreading wave packets 
(see \cite{Cherroret2021} for a review).

Recently, yet another unexpected manifestation of Anderson localization, dubbed quantum boomerang effect (QBE), has been discovered \cite{Prat2019}. The QBE corresponds to a back-and-forth motion of the mean position of a quantum wave packet launched with a finite velocity in a given direction in a random potential. In one dimension, for instance, if the quantum particle is launched to the right, it will first move to the right over a distance of the order of the mean free path, then make a U-turn and eventually return to its starting point at long time. This phenomenon was also shown to exist in higher-dimensional random or pseudo-random systems \cite{Prat2019, Tessieri2021}, as well as in kicked-rotor models \cite{Tessieri2021}, where it was recently demonstrated experimentally \cite{Sajjad2021}.
While originally described in time-reversal-invariant (TRI) systems, recently the QBE was  also shown to exist in systems without time-reversal symmetry \cite{Janarek2022, Noronha22, Macri22}. In \cite{Janarek2022}, in addition, the QBE was characterized in the presence of a spin-orbit coupling mechanism inducing left-right asymmetric scattering between different spin states. This is also the scenario addressed in the present paper.

At a theoretical level, describing the temporal dynamics of quantum wave packets in the presence of disorder is a challenging task \cite{Sanchez-Palencia2007, Skipetrov2008, Shapiro2012}. In one dimension, however, a very powerful analytical approach known as the Berezinskii diagrammatic technique has been developed \cite{Berezinskii1996}. Originally, this method was successfully used for calculating the  ac conductivity of electronic conductors in the localization regime or the long-time  density distribution of spreading wave packets \cite{Berezinskii1996, Gogolin1976, Wellens2016}, the predictions being exact in the limit of weak disorder. More recently, it also allowed to describe the QBE in TRI systems \cite{Prat2019} and, in the context of electron scattering, was extended to account for the presence of spin-orbit coupling \cite{Suleymanli2023}. 

In this paper, we extend the Berezinskii diagrammatic technique to TRI-broken spin-dependent systems in which a spin-orbit coupling term induces asymmetric scattering, as recently realized experimentally with cold atoms \cite{Lin2011, Hamner2015}. This formalism is developed in sections \ref{Sec:principles} and \ref{sec:diagrammatic_approach}. In Sec. \ref{sec:qbe} we then apply the method to the calculation of a specific observable, the mean position of a quantum-mechanical wave packet launched in a random potential with finite velocity. This provides a thorough theoretical description of the QBE in spin-orbit coupled systems with asymmetric scattering, complementing results obtained in the recent work \cite{Janarek2022}. 
Generally speaking, the formalism presented in this paper provides a practical analytical tool to characterize the dynamics of spinor wave packets in disordered systems with TRI-broken symmetry.

\section{Principles of the Berezinskii technique}
\label{Sec:principles}

We start by recalling the main ideas of the original Berezinskii diagrammatic technique used to compute the time-dependent transport properties of one-dimensional disordered systems.  The starting point is the single-particle Hamiltonian
\begin{equation}
\label{eq:generalH}
    H = H_0 + V(x),
\end{equation}
where $V(x)$ is a random (disorder) potential and $H_0$ is the disorder-free part of the Hamiltonian (e.g., $H_0 = p^2/2m$). We suppose that the random potential has a vanishing mean, $\overline{V(x)} = 0$, and follows a Gaussian statistics characterized by the two-point correlation function $\smash{\overline{V(x)V(x')} = \eta C(x'-x)}$, where $\eta$ is called the disorder strength. Symbol $\overline{(\ldots)}$ here denotes averaging over different disorder realizations. 
The function $C(x'-x)$ quantifies the range of the spatial correlation of the disorder.
In the whole paper, we 
restrict ourselves to a delta correlated potential, i.e., $C(x'-x) = \delta(x'-x)$.

In this paper, we aim at describing the time evolution of quantum-mechanical wave packets governed by an Hamiltonian of the type of Eq. (\ref{eq:generalH}). In the localization problem, this evolution is characterized by considering the disorder average of observables that depend quadratically on the wave function,  such as the density $n(x,t)$ or the mean position $\langle x(t)\rangle=\int dx\ x\ n(x,t)$ of the wave packet. These observables, by definition, can be expressed in terms of the disorder-averaged correlator $\overline{G^R(x,x',\epsilon)G^A(x'',x,\epsilon-\hbar\omega)}$ \cite{Akkermans2007book, Muller2011}, where $G^{R/A}(x,x',\epsilon) \equiv \bra{x} \left(\epsilon - H \pm i0^{+} \right)^{-1}\ket{x'}$ are the single-realization,  retarded and advanced Green's functions at energy $\epsilon$ associated with Hamiltonian (\ref{eq:generalH}). The energy difference $\hbar\omega$ introduced in the correlator allows us to capture the time dependence of observables after an inverse Fourier transform. The precise connection between $\langle x(t)\rangle$ and the Green's function correlator, for instance,  will be given in Sec. \ref{sec:diagrammatic_approach}. 

Both Green's functions $G^{R/A}$ that appear in the correlator may be computed in a perturbative fashion using the Born expansion \cite{Akkermans2007book}: 
\begin{equation}
\label{eq:born_expansion}
\begin{split}
    &G^{R/A}(x, x', \epsilon) = G^{R/A}_0(x, x', \epsilon)  \\
    &+\int\diff{x_1}G^{R/A}_0(x, x_1, \epsilon)V(x_1)G^{R/A}_0(x_1, x', \epsilon) \\ &+\int\diff{x_1}\diff{x_2}G^{R/A}_0(x, x_1, \epsilon) V(x_1)G^{R/A}_0(x_1, x_2, \epsilon) \\&\times V(x_2)G^{R/A}_0(x_2, x', \epsilon) + \ldots,
\end{split}
\end{equation}
where 
\begin{equation}\label{eq:greens_definition}
    G_0^{R/A}(x,x',\epsilon) \equiv \bra{x} \left(\epsilon - H_0 \pm i0^{+} \right)^{-1}\ket{x'}
\end{equation}
are the retarded and advanced Green's functions associated with the free part of the Hamiltonian. Physically, the expansion (\ref{eq:born_expansion}) describes a multiple scattering sequence involving scattering events on the random potential at points $x_1,\ x_2,\ \ldots$

\begin{figure}
    \centering
    \includegraphics{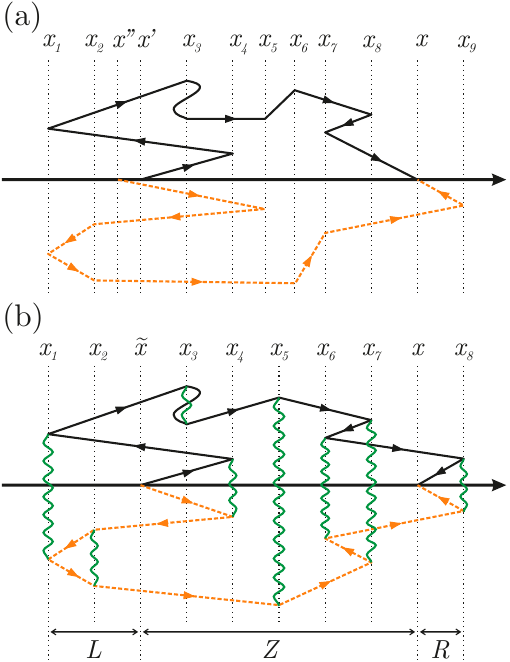}
    \caption{(a) Example of diagram involved in the product $G^R(x,x',\epsilon)G^A(x'',x,\epsilon-\hbar\omega)$ computed from the Born expansion (\ref{eq:born_expansion}). Solid lines represent free  retarded Green's functions $G^R_0$, and dashed lines free advanced  Green's functions $G^A_0$. Scattering events occur at points $x_i$. 
    (b) Example of non-vanishing contribution to the correlator $\overline{G^R(x,\tilde{x},\epsilon)G^A(\tilde{x},x,\epsilon-\hbar\omega)}$. The wavy lines refer to a two-point correlation function of the random potential.
    At each scattering event, one finds a vertex belonging to the set presented in Fig~\ref{fig:scattering_vertices}.
    The correlation diagram can be divided into three blocks $L$, $Z$, and $R$, separated by the points $\tilde{x}$ and $x$.}
    \label{fig:contributing_diagram}
\end{figure}

The average product $\overline{G^R(x,x',\epsilon)G^A(x'',x,\epsilon-\hbar\omega)}$ includes all possible correlations between two multiple scattering paths starting at the initial points $x'$ and $x''$, respectively, and both ending at the final point $x$. 
The starting point of the Berezinskii technique is to take advantage of the one-dimensional geometry, which enables us to order the  scattering events on a line:
\begin{equation}\label{eq:scattering_ordering}
    -\infty<x_1\leq \ldots \leq  x' \leq \ldots  \leq x \leq \ldots \leq x_i < \infty.
\end{equation}
Thanks to this ordering, each contribution to the product $G^R G^A$ may be represented by a diagram, like the one shown in Fig.~\ref{fig:contributing_diagram}(a) \cite{Berezinskii1996}, which combines a retarded (solid lines) and an advanced (dashed lines) multiple scattering sequence, respectively unfolded in the upper and lower parts of the diagram. The scattering events occur at the points $x_i$. In the example of Fig.~\ref{fig:contributing_diagram}(a), the upper sequence involves  8 scattering events (twice at point $x_3$), and the lower one  7 scatterings events.

The diagrams effectively contributing to the Green's function correlator $\overline{G^R G^A}$ do not have arbitrary shapes. Indeed, because of the assumed Gaussian statistics and the corresponding Wick's theorem, only diagrams whose scattering events can all be paired
appear. For instance, the diagram in Fig.~\ref{fig:contributing_diagram}(a) vanishes upon averaging because some scattering events cannot be paired.
Pairing scattering events at different points would  occur in the case of a weakly correlated disorder (for which the correlation length is smaller than the mean free path), a problem previously addressed in \cite{Rashba1976, Gogolin1976, Wellens2016}.

The second important approximation of the Berezinskii technique is to assume that among all possible diagrams contributing to the correlator, only those for which the phase factors induced by free-particle Green functions exactly compensate each other when $\omega\to0$ matter. This approximation, which holds true in the regime of ``weak disorder'' (see Sec. \ref{Sec:vertices}), amounts to imposing that there is exactly the same number of retarded and advanced Green's function in between any two successive scattering events $x_i$ and $x_{i+1}$. In turn, this 
yields restrictions on the possible \emph{scattering vertices}, which are building blocks of the diagrams: aside the trivial constraint that the solid/dashed lines have to be continuous, all vertices have to be phaseless. 

With these conditions implemented, the Berezinskii diagrammatic technique provides a strategy to exactly sum all possible diagrams with nonzero contributions, as we now detail for the case of a spin-orbit Hamiltonian $H_0$. 


\section{Diagrammatic approach without time-reversal symmetry}
\label{sec:diagrammatic_approach}

\subsection{Free Hamiltonian and Green's function}
In this work, we extend the standard Berezinskii technique to a one-dimensional spin system with spin-orbit coupling and Zeeman splitting breaking all anti-unitary time-reversal symmetries \cite{Lin2011, Hamner2015, Janarek2022}. The corresponding disorder-free Hamiltonian reads:
\begin{equation}\label{eq:h0}
    H_0 = \frac{\hbar^2 k^2}{2m} + \gamma \hbar k \sigma_z + \frac{\hbar \delta}{2}\sigma_z + \frac{\hbar \Omega}{2}\sigma_x,
\end{equation}
where  $\sigma_i$ are the usual Pauli matrices. The Hilbert space is spanned by two-dimensional complex-valued spinors. $\gamma$ is the strength of the spin-orbit coupling, $\Omega$ is the Rabi frequency and $\delta$ the detuning. 
Diagonalization of the Hamiltonian $H_0$ yields two energy bands denoted by $\pm$ with corresponding energies
\begin{equation}
    E_\pm(k) = \frac{\hbar^2 k^2}{2m} \pm \frac{\hbar}{2}\sqrt{\left(2\gamma k + \delta \right)^2 + \Omega^2}.
\end{equation}
Due to this band structure, for a given energy $\epsilon$ the Hamiltonian hosts either 2 or 4 possible eigenstates. From now on, we focus on the case where only two eigenstates are involved, which corresponds to a dynamics operating at energies belonging to the lower band only \cite{Janarek2022}. We denote by $k_\pm$ the momenta of these two states, and by $v_\pm = \frac{1}{\hbar}\left|dE_-(k_\pm)/dk\right|$ the associated velocities.
As compared to the standard single-particle Hamiltonian $\tilde{H}_0 = p^2/2m$, it should be noted that the two involved momenta are not just of opposite sign, i.e., $k_- \neq - k_+$ (and, correspondingly, $v_- \neq v_+$). The left-right symmetry is therefore broken which, as will be seen below, constitutes the most significant difference as compared to the usual Berezinskii approach. In \cite{Suleymanli2023}, a much simpler situation was studied, where only the spin-orbit interaction is present (i.e. $\delta=\Omega=0$); in such a case, the dispersion relation is symmetric with respect to $k\to -k$, so that $v_-=v_+$, and the extension of the Berezinskii technique is rather easy. In contrast, the calculations presented in the present paper are more general and valid when (generalized) TRI is broken.
 
In the diagrammatic treatment of disorder scattering introduced in the previous section, a fundamental ingredient is the free Green's function (\ref{eq:greens_definition}), which we need to evaluate for the Hamiltonian (\ref{eq:h0}). To this aim, we use the definition
\begin{equation}
\label{eq:G0def}
    G_0^R(x,x',\epsilon)\equiv \int_{-\infty}^\infty\frac{dk}{2\pi}\frac{e^{ik(x-x')}}{\epsilon-E_-(k)+i0^+}.
\end{equation}
A careful calculation of the integral in momentum space provides us with 
\begin{equation}
\label{eq:so_green_function}
    G_0^R(x, x', \epsilon) = 
    \begin{cases}
        -\frac{i}{\hbar v_+}e^{ik_+(x-x')}, \quad x - x' > 0 \\
        -\frac{i}{2\hbar}\left(\frac{1}{v_+} + \frac{1}{v_-}\right), \quad x = x' \\
        -\frac{i}{\hbar v_-}e^{ik_-(x-x')}, \quad x - x' < 0
    \end{cases}
\end{equation}
where, in particular, the diagonal value $ G_0^R(x, x, \epsilon$) is obtained by properly accounting for all the real and complex poles in the denominator of Eq. (\ref{eq:G0def}). Note that, strictly speaking, when $x-x'\ne0$ these expressions only hold at distances $|x-x'|$ larger than the de Broglie wavelength $2\pi/|k_\pm|$. 
This knowledge, however, is sufficient 
within the weak disorder limit (\ref{eq:weak_disorder}) where the Berezinskii approach operates. 
Because of translation invariance, the free Green's function $G_0^R(x,x',\epsilon) = G_0^R(x-x',\epsilon)$. Its Fourier transform is therefore diagonal in momentum space, with the diagonal value defined as $G_0^R(k, \epsilon) = \int\diff{r} e^{-ikr} G_0^R(r, \epsilon)$, where $r=x-x'$. Note that with this definition, Eq. (\ref{eq:so_green_function}) implies that $G^{R/A}(k,\epsilon)\neq G^{R/A}(-k,\epsilon)$,  contrary to TRI systems.
From Eq. (\ref{eq:so_green_function}), finally, the advanced Green's function follows from Hermitian conjugation, $G_0^A(x,x',\epsilon)=[G_0^R(x',x,\epsilon)]^*$. 

In the following, we will also need the energy-shifted Green's function $G_0^A(x', x, \epsilon - \hbar \omega)$, where $\omega \ll \epsilon/\hbar$. To evaluate this object, we use the Taylor expansions $k_\pm(\epsilon-\hbar\omega)\approx k_\pm\mp\omega/v_\pm$, so that:
\begin{eqnarray}
\label{eq:gA_omega}
    G_0^A&&\!\!\!\!\!\!\!\!(x', x, \epsilon-\hbar\omega) =\nonumber \\
   && \begin{cases}
        \frac{i}{\hbar v_+}e^{-i(k_+-\omega/v_+)(x-x')}, \quad x - x' > 0 \\
        \frac{i}{2\hbar}\left(\frac{1}{v_+} + \frac{1}{v_-}\right), \quad x = x'\\
        \frac{i}{\hbar v_-}e^{-i(k_-+\omega/v_-)(x-x')}, \quad x - x' < 0
    \end{cases}
\end{eqnarray}

\subsection{Mean free times}
\label{Sec:mft}

Before constructing the diagrammatic approach based on the Hamiltonian $H_0+V(x)$, let us introduce 
a few important scattering parameters that will be used in the following. The central one is the concept of scattering mean free time, which gives the average time scale between two consecutive scattering events. In the present case, however, two different mean free paths can be defined due to the left-right asymmetry. 
To find them, let us denote  by $|\pm\rangle=|k_\pm\rangle\otimes|\chi_\pm\rangle$ the two eigenstes of $H_0$, where $\ket{\chi_+}$ and $\ket{\chi_-}$ are the spin state components associated with the wave numbers $k_+$ and $k_-$, respectively.  This leads us to define $\tau_+$ and $\tau_-$, the scattering mean free times for the processes $|+\rangle\to |-\rangle$ and $|-\rangle\to |+\rangle$, respectively. At weak disorder they can be evaluated from the Fermi golden rule
\begin{equation}
\label{eq:taudef}
    \frac{1}{\tau_\pm}=\frac{2\pi}{\hbar} \,
    \overline{|\langle \mp|V|\pm\rangle|^2}\, \rho(E_-(k_\mp)),
\end{equation}
where $\rho(E_-(k_\mp))$ is the density of states evaluated at the energy $E_-(k_\mp)$ of the final state. Using that the disorder is uncorrelated, i.e., $\overline{V(x')V(x)}=\eta\delta(x'-x)$ (see Sec. \ref{Sec:principles}), we infer:
\begin{equation}
\label{eq:tau_pm}
    \tau_\pm = \frac{\hbar^2 v_\mp}{2\eta\kappa},
\end{equation}
where $\kappa \equiv \left|\bra{\chi_+}\ket{\chi_-}\right|^2$ is the overlap factor of the two spin states. In the following we will be also led to use the mean free time associated with the weighted sum of the two scattering processes:
\begin{equation}
\label{eq:taudef}
\frac{1}{\tau}=\frac{1}{2}\left(\frac{1}{\tau_+}+\frac{1}{\tau_-}\right),
\end{equation}
which turns out to be the relevant time scale governing the boomerang effect, as will be shown in Sec. \ref{sec:qbe}. 
Note that the validity of the Fermi golden rule used above is only guaranteed in the weak disorder limit described by Eq. (\ref{eq:weak_disorder}) below.

\subsection{Vertices}
\label{Sec:vertices}
\begin{figure}
    \centering
    \includegraphics[width=0.9\linewidth]{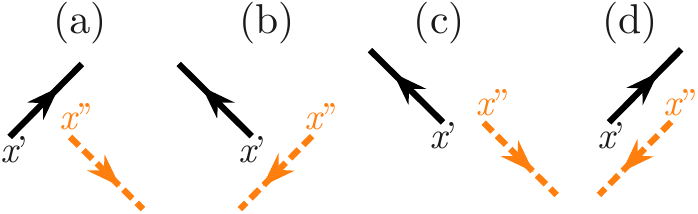}
    \caption{List of all possible initial vertices. In the limit $\omega\to0$ they correspond to the factors
    (a) $(\hbar v_+)^{-1} e^{ik_+(x''-x')}$,
    (b) $(\hbar v_-)^{-1} e^{ik_-(x''-x')}$, (c) $(\hbar^2v_-v_+)^{-1/2}e^{i(k_+x''-k_-x')}$, and (d) $(\hbar^2v_-v_+)^{-1/2}e^{i(k_-x''-k_+x')}$.}
    \label{fig:init_vertices}
\end{figure}
At the core of the Berezinskii diagrammatic technique is the idea to transfer the propagating factors from the free Green's functions to the scattering events, called vertices. For example, assuming $x_i > x_j$, the free Green's function can be split as
\begin{equation}
\label{eq:greens_transfer_to_points}
    G_0^R(x_i, x_j, \epsilon) = \sqrt{-\frac{i}{\hbar v_+}}e^{ik_+ x_i}\sqrt{-\frac{i}{\hbar v_+}}e^{-ik_+ x_j},
\end{equation}
where we formally associate the weights and exponential factors to the vertices at points $x_i$ and $x_j$. The difference between TRI system and the TRI-broken case is that these factors depend on the direction of propagation. For example, in the TRI system the opposite case $x_j > x_i$ would result in just a change of sign of the phase factors in Eq.~(\ref{eq:greens_transfer_to_points}), whereas in the system with broken TRI also the velocities change. \\

\paragraph*{Initial vertices.}
We start by selecting the relevant initial vertices effectively contributing to the correlator $\overline{G^R(x,x',\epsilon)G^A(x'',x,\epsilon-\hbar\omega)}$. In general, scattering paths may start from any of the 4 vertices shown with their weights in Fig.~\ref{fig:init_vertices}. The vertices with advanced and retarded lines starting into opposite directions, i.e., vertices (c) and (d), carry exponential factors with phases $i(k_\pm x'' - k_\mp x')$. Upon integration over the starting points $x'$ and $x''$ [cf. Eq.~(\ref{eq:com_freq1}) in Sec.~\ref{sec:qbe}], they typically yield negligible contributions. Thus, we can restrict the analysis  to only two classes of initial vertices: (a) and (b). These classes, in turn, correspond to two different types of initial states for the dynamics, (a) with positive ($v_+$) and (b) with negative ($v_-$) initial velocity. 

A second simplification is based on the assumption that no scattering happens between the initial points $x'$ and $x''$ \cite{Prat2017a, Prat2019}. This invites us to introduce the Wigner variables $r = x' - x''$ and $\tilde{x} = (x' + x'')/2$. In the limit $\omega \to 0$, vertices (a) and (b) are thus approximated by their counterparts starting from a single point $\tilde{x}$. At the level of Green's functions, this simplification reads \cite{Wellens2016}:
\begin{equation}
\begin{split}\label{eq:initial_vertex_simplification}
    \overline{G^R(x,x',\epsilon)G^A(x'',x,\epsilon-\hbar\omega)} \approx \\
    e^{-ik_\epsilon r}\overline{G^R(x,\tilde{x},\epsilon)G^A(\tilde{x},x,\epsilon-\hbar\omega)}
\end{split}
\end{equation}
where $k_\epsilon$ is the wave number  satisfying the dispersion relation $\epsilon=E_-(k_\epsilon)$. \\

\paragraph*{Phaseless scattering vertices}

\begin{figure}
    \centering
    \includegraphics[width=0.7\linewidth]{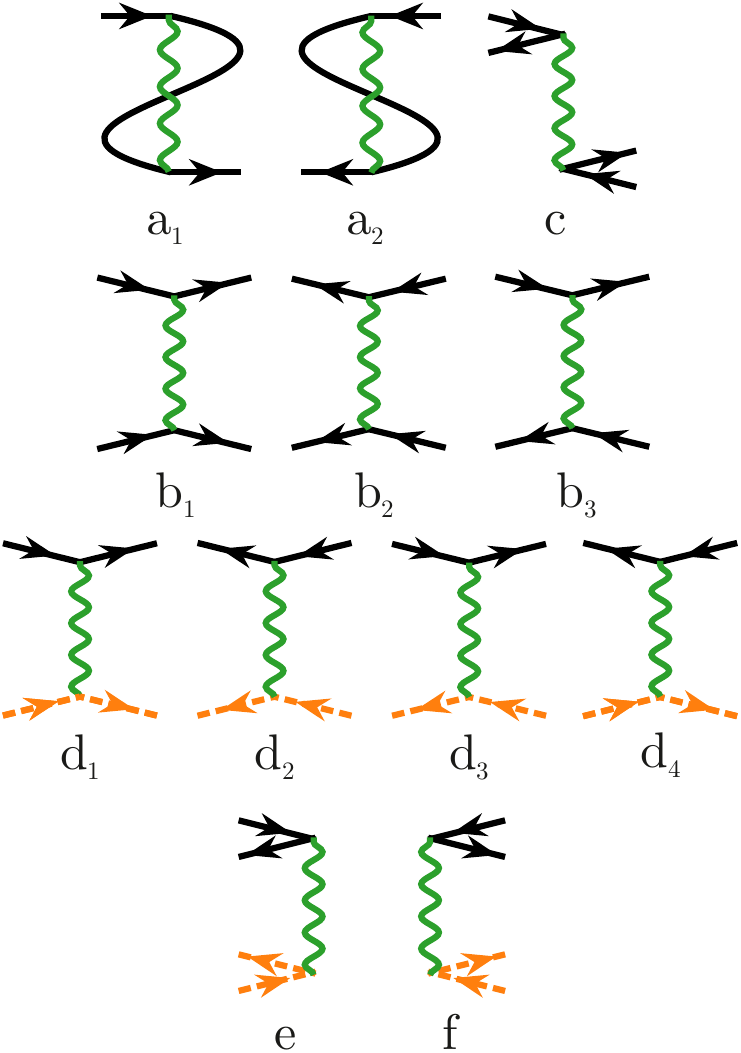}
    \caption{All possible phaseless scattering vertices to be considered in the Berezinskii technique. Vertices from a, b and c families have dashed-lines counterparts. The weights associated to the vertices are indicated in Eq.~(\ref{eq:vertices_weights}).}
    \label{fig:scattering_vertices}
\end{figure}

Our ultimate goal is to sum all significant contributions to the product of Green's functions $\overline{G^R G^A}$. This formidable task is, in general, out of reach except in the so-called weak-disorder limit
\begin{equation}
\label{eq:weak_disorder}
    k_\epsilon \ell\gg1,
\end{equation}
where $\ell=\tau_+v_+=\tau_- v_-$ is the scattering mean free path. Under this condition, only a limited set of scattering vertices that do not accumulate any phase and, as such, are not vanishingly small upon disorder averaging, should be considered when constructing correlation diagrams. 
The procedure to identify this set is detailed in Appendix \ref{Appendix:vertices} for clarity. It yields four families of vertices that are listed in Fig. \ref{fig:scattering_vertices}. One can easily check that the phase associated with each vertex is zero. For instance, the vertex a$_1$ originates from a factor of the type $ \eta G_0^R(x_i,x_{i-1})G_0^R(x_i,x_i) G_0^R(x_{i+1},x_i)$ in the disorder average of the Born expansion (\ref{eq:born_expansion}). With the help of the 
splitting procedure (\ref{eq:greens_transfer_to_points}) and of Eq. (\ref{eq:so_green_function}), this corresponds to a vertex weight
\begin{equation}
    \eta \sqrt{\frac{-i}{\hbar v_+}}e^{ik_+x_i}\times \frac{-i}{2\hbar}\Big(\frac{1}{v_+}+\frac{1}{v_-}\Big)\times \sqrt{\frac{-i}{\hbar v_+}}e^{-ik_+x_i},
\end{equation}
whose phase is indeed zero. In turn, the weights of all  phaseless scattering vertices are 
\begin{align}\label{eq:vertices_weights}
    &\text{a\textsubscript{1/2}: } -\frac{\eta}{2\hbar^2 v_\pm}\left(\frac{1}{v_+} + \frac{1}{v_-}\right),\nonumber\\
    &\text{b\textsubscript{1/2}: } -\frac{\eta}{(\hbar v_\pm)^2}, \quad \text{b\textsubscript{3}: } -\frac{\eta}{(\hbar v_+)(\hbar v_-)} \nonumber\\
    &\text{c: }  - \frac{\eta\kappa}{(\hbar v_+)(\hbar v_-)} \\
   & \text{d\textsubscript{1/2}: } \frac{\eta}{(\hbar v_\pm)^2}, \quad
    \text{d\textsubscript{3/4}: } \frac{\eta}{(\hbar v_+)(\hbar v_-)}, \nonumber\\
   & \text{e: } \frac{\eta\kappa}{(\hbar v_+)(\hbar v_-)}\exp\left[i\omega x \left(\frac{1}{v_+} + \frac{1}{v_-}\right)\right] \nonumber\\
    &\text{f: } \frac{\eta\kappa}{(\hbar v_+)(\hbar v_-)}\exp\left[-i\omega x \left(\frac{1}{v_+} + \frac{1}{v_-}\right)\right]\nonumber
\end{align}
Notice that among all diagrams in Fig. \ref{fig:scattering_vertices}, the 
 vertices families c, e, and f involve a ``backscattering event'' in both the retarded and advanced parts. This implies that, in the spin system described by Hamiltonian~(\ref{eq:h0}), the associated weights include the spin-state overlap factor $\kappa = \left|\bra{\chi_+}\ket{\chi_-}\right|^2$. 

\subsection{Correlation diagrams}\label{sec:equations_for_diagrams}

Knowing all possible phaseless scattering vertices relevant to our problem, we now wish to write down the equations describing the  diagrams contributing to the correlator $\smash{\overline{G^R G^A}}$. An example of such a correlation diagram 
is shown in Fig.~\ref{fig:contributing_diagram}(b). Its generic structure can be divided into three left, right and central blocks denoted by $L$, $R$, and $Z$, as illustrated in Fig.~\ref{fig:contributing_diagram}(b). These different blocks are characterized by their total number of incoming and outgoing solid (retarded) and dashed (advanced) lines. 

We first consider the left blocks $L$. Because the scattering vertices change the number of lines by at most 2, these blocks  always have the same even number $2m'$ (with $m'$ an integer) of retarded and advanced lines attached. For instance, $L$ in Fig.~\ref{fig:contributing_diagram}(b) has $m' = 1$. With this property in mind, let us denote by $L_{m'}(\tilde{x})$ the sum of contributions from all $L$ blocks that have their right boundary at point $\tilde{x}$ with $2m'$ lines.
To calculate $L_{m'}(\tilde{x})$, we consider how it changes with an infinitesimal change of the boundary position, $\tilde{x}\to\tilde{x}+\delta x$, by counting all possibilities of adding new scattering vertices to $L_{m'}(\tilde{x})$. This  counting in detailed in Appendix \ref{Appendix:Lmblock} for clarity. Taking the limit $\delta x\to 0 $, it yields the following differential equation \cite{Janarek2021}:
\begin{align}
    \label{eq:l_m_full}
        &\dfrac{L_{m'}}{\tilde{x}} = -\frac{2m'\eta}{\hbar^2 v_+v_-}L_{m'}\left[1 + (m'-1)\kappa\right] +\frac{m'^2 \eta \kappa}{\hbar^2 v_+ v_-}\times\nonumber\\
        &\left[L_{m'+1}e^{i\omega \tilde{x}(\frac{1}{v_+} + \frac{1}{v_-})}\!+\!L_{m'-1}e^{-i\omega \tilde{x}(\frac{1}{v_+} + \frac{1}{v_-})}\right]\!.
\end{align}
This equation is solved by an Ansatz $L_{m'}(\tilde{x}) = \mathcal{L}_{m'}\exp\left[-im'\omega \tilde{x}({1}/{v_+} + {1}/{v_-})\right]$, which leads to an iterative equation for $\mathcal{L}_{m'}$:
\begin{equation}\label{eq:l_m_algebraic}
    s \mathcal{L}_{m'} + m'(\mathcal{L}_{m'+1} - \mathcal{L}_{m'-1} + 2\mathcal{L}_{m'}) = 0,
\end{equation}
where $s = 2  - {2}/{\kappa} + i\nu$ with $\nu=\omega (v_+ + v_-)\hbar^2/\kappa \eta$. The explicit solution of Eq.~(\ref{eq:l_m_algebraic}) is
\begin{equation}
\label{eq:Lmfinal}
\mathcal{L}_m(s) = - s\Gamma(m+1)\Psi(m+1,2;-s),
\end{equation}
 with $\Psi(a,b;z)$ the confluent hypergeometric function of the second kind. Note that in the usual case of spinless TRI systems, $\mathcal{L}_m$ satisfies a similar equation as Eq.~(\ref{eq:l_m_algebraic}), but with $s_\text{TRI} = 2i\omega v/\eta$, where $v$ is the velocity of the state at energy $\epsilon$ \cite{Berezinskii1996}. The main difference is that $s_\text{TRI}$ is fully imaginary, while in our case $s$ has a finite real part. 

The treatment of the right block $R$ is fully analogous. Denoting by $R_m(x)$ the sum of all right-hand blocks which have their left boundary at point $x$ with $2m$ lines (with $m$ an integer), we find that $R_m(x) = L_m(-x)$ and, with a similar Ansatz, $\mathcal{R}_m = \mathcal{L}_m$.

Let us finally consider the central block $Z$. As compared to $L$ and $R$, this block has one additional line which connects points $\tilde{x}$ and $x$, i.e., for left and right block having $2m'$ and $2m$ retarded and advanced lines attached, the central block $Z_{m',m}$ connecting  them has  $2m'+1$ lines at its left boundary and $2m+1$ lines at its right boundary. For instance, the diagram in Fig.~\ref{fig:contributing_diagram}(b) has $2m'+1=3$ and $2m+1=1$. 
\begin{figure}
    \centering
    \includegraphics[width=0.7\linewidth]{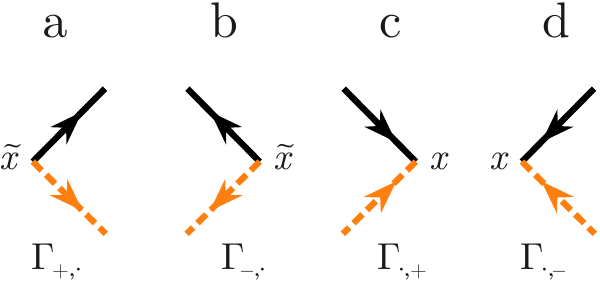}
    \caption{List of possible initial and final phaseless vertices. They are associated with the following weights (for $\omega\to 0$): (a) $\Gamma_{+,\cdot} = (\hbar v_+)^{-1}$, (b) $\Gamma_{-,\cdot} = (\hbar v_-)^{-1}$, (c) $\Gamma_{\cdot, +} = (\hbar v_+)^{-1}$, and (d) $\Gamma_{\cdot, -} = (\hbar v_-)^{-1}$.}
    \label{fig:initial_final_vertices}
\end{figure}
To derive a differential equation for $Z_{m',m}(\tilde{x},x)$, we have to make an assumption on the direction of the extra line. Its type depends on the sign of $x-\tilde{x}$ and introduces a kind of asymmetry because our problem differentiates left and right directions. Here, we assume $\tilde{x} - x < 0$, that is the additional line is going from left to right, like in Fig.~\ref{fig:contributing_diagram}(b).
The total derivative of $Z_{m',m}(\tilde{x},x)$ with respect to $x$ includes the contributions from scattering vertices but also it has to include the derivative of the final vertex. These final vertices are analyzed analogously to the initial vertices. Out of four possibilities only two are phaseless, and thus contribute to the final sum of diagrams. They correspond to vertices with lines incoming only from a single direction, i.e., both from left or both from right. The list of all phaseless initial and final vertices is summarized in Fig.~\ref{fig:initial_final_vertices}, together with their corresponding weights, denoted by $\Gamma_{\pm,.}$ and $\Gamma_{.,\pm}$ for initial and final vertices, respectively.

Computing the total derivative of the central block at the final point $x$,  assuming $\tilde{x} < x$, we find:
\begin{align}\label{eq:Z_m}
    &\dfrac{ Z_{m',m}(\tilde{x},x)}{x} =\nonumber \\
   & \pm \frac{i\omega}{v_\pm}Z_{m',m}(\tilde{x},x) - \frac{\eta}{\hbar^2 v_+ v_-} \left(2m^2\kappa + 2m + 1\right)Z_{m',m}(\tilde{x},x) \nonumber\\
    &+\frac{\eta \kappa}{\hbar^2 v_+v_-}\left[(m+1)^2 Z_{m',m+1}(\tilde{x},x)e^{i\omega x(\frac{1}{v_+} + \frac{1}{v_-})} \right. \nonumber\\
    &+\left.  m^2 Z_{m',m-1}(\tilde{x},x)e^{-i\omega x(\frac{1}{v_+} + \frac{1}{v_-})} \right].
\end{align}
The sign of the first term in the right-hand side depends on the final vertex type, i.e., $\Gamma_{\cdot, +}$ or $\Gamma_{\cdot, -}$.
It turns out, on the other hand, that this expression does not depend on the sign of $\tilde{x} - x$. Note that when $v_+ = v_-$ and $\kappa = 1$, Eqs.~(\ref{eq:l_m_full}) and (\ref{eq:Z_m}) reduce to the known spinless TRI case \cite{Berezinskii1996}. 
While we are not aware of any analytic solution for the differentio-recursive equation (\ref{eq:Z_m}), in general the direct knowledge of the full function $Z_{m',m}(\tilde{x},x)$ is not required for the computation of observables. An example of this will be given in the next section, when discussing the quantum boomerang effect.

We conclude this section by expressing the Green's function correlator in Eq. (\ref{eq:initial_vertex_simplification}) in terms of the blocks $L$, $R$ and $Z$ described above. For $\tilde{x}<x$, and if we suppose that the initial wave function only populates the state with initial velocity $v_+$ (this is the practical case that will be considered in Sec.  \ref{sec:qbe}), the correlator 
\begin{align}
\label{eq:GRG1final}
\overline{G^R(x,\tilde{x},\epsilon)G^A(\tilde{x},x,\epsilon\!-\!\hbar\omega)}\!=\!\frac{1}{\hbar^2 v_+^2}\Gamma_{+,+}^{\tilde{x}<x}\!+\!\frac{1}{\hbar^2 v_+v_-}\Gamma_{+,-}^{\tilde{x}<x}
\end{align}
is the sum of two contributions corresponding to the two possible final vertices c and d in Fig. \ref{fig:initial_final_vertices}, with
\begin{align}
    &\Gamma_{+,+}^{\tilde{x}<x}=
    \sum_{m,m'=0}^\infty
    \mathcal{L}_{m'}(\tilde{x})Z_{m',m}(\tilde{x},x)\mathcal{R}_m(x)\\
    &\Gamma_{+,-}^{\tilde{x}<x}=
    \sum_{m,m'=0}^\infty
    \mathcal{L}_{m'}(\tilde{x})Z_{m',m}(\tilde{x},x)\mathcal{R}_{m+1}(x).
\end{align}
In the opposite case $\tilde{x}>x$, finally, Eq. (\ref{eq:GRG1final}) still holds but with $\Gamma_{+,\pm}^{\tilde{x}<x}$ changed to
\begin{align}
    &\Gamma_{+,+}^{\tilde{x}>x}=
    \sum_{m,m'=0}^\infty
    \mathcal{L}_{m'+1}(x)Z_{m',m}(x,\tilde{x})\mathcal{R}_{m+1}(\tilde{x})\\
   & \Gamma_{+,-}^{\tilde{x}>x}=
    \sum_{m,m'=0}^\infty
    \mathcal{L}_{m'}(x)Z_{m',m}(x,\tilde{x})\mathcal{R}_{m+1}(\tilde{x}).
    \label{eq:gammapm}
\end{align}
Together with the solution of Eq. (\ref{eq:Z_m}), Eqs. (\ref{eq:GRG1final}--\ref{eq:gammapm}) constitute the final solution of the localization problem. In the next section, we apply this formalism to access the time evolution of a particular observable, the mean position of wave packets, featuring the quantum boomerang effect.

\section{Quantum boomerang effect without time-reversal symmetry}
\label{sec:qbe}

We now apply the above formalism to the theoretical  description of a concrete problem, the quantum boomerang effect (QBE). We recall that the QBE describes a back-and-forth motion of the mean position of a quantum particle  launched with nonzero initial velocity in a disordered potential. Here we describe this phenomenon based on the TRI-broken Hamiltonian $H_0+V$, with the free part $H_0$ defined by Eq. (\ref{eq:h0}).

\subsection{Mean position}

To describe the QBE within the Berezinskii technique, we consider for definiteness a wave packet initially launched in a disordered potential with the mean eigen wave number $k_+$ of the Hamiltonian (\ref{eq:h0}) in the corresponding spin state $|\chi_+\rangle$. We denote by $\epsilon_0=E_-(k_+)$ the associated energy. 
We thus write the initial wave function as
\begin{equation}
\label{eq:initialWF}
    \Psi_0(x)=\frac{1}{(\pi\sigma^2)^{1/4}}
    \exp\left(-\frac{x^2}{2\sigma^2}+i k_+x\right)
    |\chi_+\rangle,
\end{equation}
where $\sigma$ is the wave-packet width. 
As explained in the previous sections, the ensuing dynamics of this state in the disorder gives rise to a coupling with the backward-propagating state of wave number $k_-$ and spin component $|\chi_-\rangle$. 

By definition, the disorder-average mean position is
\begin{equation}
        \langle x(t) \rangle \equiv \int \diff{x} x \overline{|\psi(x,t)|^2}.
\end{equation}
Using that $\psi(x,t) = \int\diff{x'}G^R(x,x',t)\Psi_0(x')$,
we can relate its Fourier transform $\langle x(\omega)\rangle=\int dt e^{i\omega t} \langle x(t)\rangle$ to the Green's function correlator as
\begin{equation}
\begin{split}\label{eq:com_freq1}
   \langle x (\omega)\rangle  &= \frac{1}{2\pi\hbar} \int \diff{x}\diff{x'}\diff{x''}\diff{\epsilon} \Psi_0(x')\Psi^*_0(x'')\times \\ &x\, \overline{G^R(x, x', \epsilon)G^A(x'', x, \epsilon - \hbar \omega)},
\end{split}
\end{equation}
where we expressed the retarded and advanced Green's functions in the Fourier domain.
To simplify this expression, we make use of  Eq.~(\ref{eq:initial_vertex_simplification}), which leads to 
\begin{equation}
    \begin{split}
        \langle x(\omega)\rangle\! =\!\! \int\diff{x}\diff{\tilde{x}}\diff{\epsilon} x\, W(\tilde{x}, k_\epsilon) 
\overline{G^R(x,\tilde{x},\epsilon)G^A(\tilde{x},x,\epsilon\!-\!\hbar\omega)}\nonumber
    \end{split}
\end{equation}
with $W$ the Wigner distribution of the initial state:
\begin{equation}
    2\pi\hbar W(\tilde{x}, k_\epsilon)
    =\int dr e^{-i k_\epsilon r}
    \Psi_0(\tilde{x}+r/2)\Psi_0^*(\tilde{x}-r/2).
\end{equation}
For an initial wave function (\ref{eq:initialWF}) of spatial width $\sigma$ much smaller than the mean free path, we find $W(\tilde{x}, k_\epsilon)\approx \hbar^{-1}\delta(\tilde{x})\delta(k_\epsilon-k_+)$, such that, eventually,
\begin{equation}\label{eq:com_freq2}
    \langle x(\omega)\rangle \!=\! v_+\!\int_{-\infty}^\infty\!\!\!\!
    \diff{\Delta x}
    \Delta x\, \overline{G^R(x,\tilde{x},\epsilon_0)G^A(\tilde{x},x,\epsilon_0\!-\!\hbar\omega)},
\end{equation}
where for convenience we replaced the integral over $x$ by an integral over $\Delta x\equiv x-\tilde{x}$, using that the integrand depends only on $x-\tilde{x}$ due to statistical translational invariance. Equation (\ref{eq:com_freq2}) directly connects the average mean position to the Green's function correlator, which we now compute using the results of the previous section.

\subsection{Time evolution of the boomerang effect}

Inserting the general Berezinskii result (\ref{eq:GRG1final}) into Eq. (\ref{eq:com_freq2}), we infer
\begin{equation}
    \langle x(\omega)\rangle=\langle x(\omega)\rangle_++\langle x(\omega)\rangle_-,
\end{equation}
where
\begin{equation}\label{eq:com_+}
        \langle x(\omega)\rangle_{+} = \frac{2\ell}{v_+}\sum_{m'} \left(\mathcal{L}_{m'} S^0_{m'} + \mathcal{L}_{m'+1}S^1_{m'}\right)
\end{equation}
is the contribution of velocities $v_+$ [technically, of the final vertex c in Fig. \ref{fig:initial_final_vertices}], and
\begin{equation}\label{eq:com_-}
        \langle x(\omega)\rangle_{-} = \frac{2\ell}{v_-}\sum_{m'} \left(\mathcal{L}_{m'} S^2_{m'} + \mathcal{L}_{m'}S^3_{m'}\right)
\end{equation}
is the contribution of velocities $v_-$ [final vertex d in Fig. \ref{fig:initial_final_vertices}]. Notice that we here introduced for convenience the mean free path $\ell=\tau_+v_+=\tau_-v_-$. In Eqs. (\ref{eq:com_+}) and (\ref{eq:com_-}), 
the two terms in the right-hand side are the contributions of $\tilde{x}<x$ and $\tilde{x}>x$, respectively, with the coefficients $\mathcal{L}_m$ defined by Eq. (\ref{eq:Lmfinal}). The quantities $S_m^i$, on the other hand, are given by spatial integrals of the block functions $\mathcal{R}_m$ and $Z_{m',m}$. For instance, we have
\begin{equation}
    \begin{split}
    S_{m'}^0=\frac{1}{2\ell}\sum_m &\int_0^\infty d\Delta x \Delta x e^{-im'\omega \tilde{x}(\frac{1}{v_+}+\frac{1}{v_-})}\\
    &\times Z_{m',m}(\tilde{x},x)
    e^{im\omega x(\frac{1}{v_+}+\frac{1}{v_-})}\mathcal{R}_m.
\end{split}
\end{equation}
To compute the coefficients $S_{m'}^0$, we perform a partial integration in the right-hand side and use Eq. (\ref{eq:Z_m}) to express the derivative $dZ_{m',m'}(\tilde{x},x)/d\tilde{x}$ in terms of $Z_{m',m}$. This provides us with the iterative equation
\begin{equation}
\label{eq:iterativeS0m}
\begin{split}
        2\ell Q_m^0&+i\nu\left(m+\frac{v_-}{v_++v_-}\right)S_m^0-2\ell \eta\beta_m S_m^0\\
    &+m^2 S_{m-1}^0+(m+1)^2 S_{m+1}^0=0
\end{split}
\end{equation}
where $\beta_m\equiv (2\kappa m^2+2m+1)/(\hbar^2v_+ v_- )$ and we remind that $\nu\equiv \omega(v_++v_-)\hbar^2/(\kappa\eta)$. The coefficient $Q_m^0$ is defined as 
\begin{equation}
    \begin{split}
        Q_{m}^0=\frac{1}{2\ell}\sum_m \int_0^\infty& d\Delta x e^{-im'\omega \tilde{x}(\frac{1}{v_+}+\frac{1}{v_-})}\\
    &\times Z_{m',m}(\tilde{x},x)
    e^{im\omega x(\frac{1}{v_+}+\frac{1}{v_-})}\mathcal{R}_m
\end{split}
\end{equation}
and is deduced from an iterative equation similar to Eq. (\ref{eq:iterativeS0m}):
\begin{equation}
\label{eq:iterativeQ0m}
\begin{split}
    \mathcal{L}_m&+i\nu\left(m+\frac{v_-}{v_++v_-}\right)Q_m^0-2\ell \eta\beta_m Q_m^0\\
    &+m^2 Q_{m-1}^0+(m+1)^2 Q_{m+1}^0=0.
\end{split}
\end{equation}
The coupled system of equations (\ref{eq:iterativeS0m}) and (\ref{eq:iterativeQ0m}) is closed, so that at a formal level it  can in principle be  solved to find the coefficients $S_m^0$ and, in turn, to compute the first sum in the right-hand side of Eq. (\ref{eq:com_+}). The calculation of the coefficients $S_m^1$, $S_m^2$ and $S_m^3$ that appear in Eqs. (\ref{eq:com_+}) and (\ref{eq:com_-}) follows the same lines. We provide the corresponding iterative equations they obey in Appendix \ref{Appendix:iterative} for the sake of completeness.

\begin{figure}
    \centering
    \includegraphics{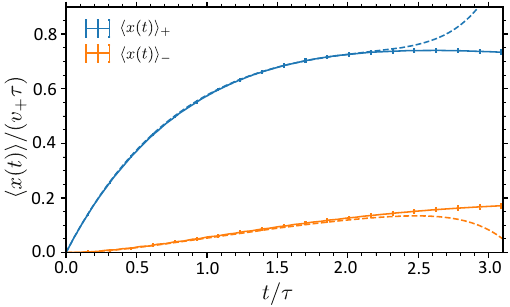}
    \caption{Numerical (solid lines with error bars) and  short-time diagrammatic solution (dashed lines) for the mean position  computed up to order $\mathcal{O}(t^{11})$ for a disordered system whose free Hamiltonian part is given by Eq.~(\ref{eq:h0}). The numerical simulations consist in a temporal propagation of the initial state (\ref{eq:initialWF}) with the Shr\"{o}dinger equation. For these simulations we 
    choose $\sigma=50$ for the wave-packet width,
     $\gamma=0.4$, $\delta=\Omega=0.4$ for the Hamiltonian parameters, and an energy $\epsilon_0 = 0$. We take $v_-/v_+=0.66$ for the ratio of velocities,  and a disorder strength $\eta=0.0049$. 
     The simulations are done using a system of length $L=10000$ with a small discretization $\Delta x= 0.2$, and numerical results are averaged over 40960 disorder realizations.}
    \label{fig:com_short_time}
\end{figure}

Instead of exactly computing all the $S_m^i$ coefficients, 
however, the mean position can be conveniently evaluated from its short-time expansion using a Pad\'e approximant \cite{Prat2017a, Janarek2021}. The short-time expansion of $\langle x(t)\rangle_+$ and $\langle x(t)\rangle_-$ is systematically obtained by inserting the series $S_m^i(\nu)=\sum_n s^i_{m,n}/(i\nu)^n$ and $Q_m^i(\nu)=\sum_n q^i_{m,n}/(i\nu)^n$ in the iteratives relations (\ref{eq:iterativeS0m}), (\ref{eq:iterativeQ0m}) and (\ref{eq:iterativeS1m})--(\ref{eq:iterativeQ3m}), and computing the $s_{m,n}^i$ and $q_{m,n}^i$ coefficients at arbitrary order in $1/\nu$. This procedure eventually yields the following short-time expansion for the mean position:
\begin{equation}
\label{eq:com_quantum_short_time}
    \begin{split}
        &\frac{\langle x(t)\rangle}{v_+\tau} =  \frac{t}{\tau} - \frac{t^2}{2\tau^2} +  \frac{t^3}{6\tau^3} \\
        & - \frac{(1+\Delta(4+\Delta(8+\Delta(4+\Delta))))t^4}{24(1+\Delta)^4 \tau^4} + \mathcal{O}(t^5)
    \end{split}
\end{equation}
where $\Delta\equiv v_-/v_+$ and $\tau$ is defined by Eq. (\ref{eq:taudef}). In Appendix \ref{Appendix:short-time}, we also provide the corresponding expansions for the partial components $\langle x(t)\rangle_+$ and $\langle x(t)\rangle_-$ that respectively describe right and left-moving particles after the last scattering event. 
In Fig. \ref{fig:com_short_time} we show a comparison between an exact numerical calculation of $\langle x(t)\rangle_\pm$ based on a temporal wave-packet propagation with the disordered Schrödinger equation (details on the numerical simulations are given in the figure caption), and the short-time expansion up to order 11 obtained by solving the Berezinskii equations as explained above.  Numerical and theoretical results are in very good agreement without any fit parameter up to $t/\tau\approx 3$. This corresponds to a finite radius of convergence in time, which is also present in the  TRI version of the quantum boomerang effect \cite{Prat2019}. This radius seems to be dependent of the ratio of velocities $\Delta$. Most importantly, as indicated by Eq. (\ref{eq:com_quantum_short_time}), the expression of $\langle x(t)\rangle$ is no longer universal because it depends on the velocities' ratio starting from the 4th order. This  is a significant difference with the TRI quantum boomerang effect, which solely depends on the dimensionless time scale $t/\tau$ at all times. The TRI solution is fully recovered when $v_+ = v_-=v$.

It is also instructive to compare the exact, quantum-mechanical short-time expansion (\ref{eq:com_quantum_short_time}) with the classical prediction of the Boltzmann equation, which discards any interference in the multiple scattering process. At a classical level, the mean position is simply given by $\langle x(t)\rangle^\text{class.} = \tau v_+(1-e^{-t/\tau})$, which is essentially the same expression as in TRI systems (see the supplemental material of \cite{Janarek2022} for details on the classical calculation). This classical result has the short-time expansion
\begin{equation}
\begin{split}
    \frac{\langle x(t) \rangle^\text{class.}}{v_+\tau} = \frac{t}{\tau} - \frac{t^2}{2\tau} + \frac{t^3}{6\tau^3} - \frac{t^4}{24\tau^4} + \mathcal{O}(t^5)
\end{split}
\end{equation}
which starts to deviate from the quantum-mechanical prediction (\ref{eq:com_quantum_short_time}) starting from the 4th order. For completeness, in Appendix \ref{Appendix:short-time} we also provides the short-time expansions for the classical components $\langle x(t)\rangle_+^\text{class.}$ and $\langle x(t) \rangle_-^\text{class.}$.
\begin{figure}
    \centering
    \includegraphics{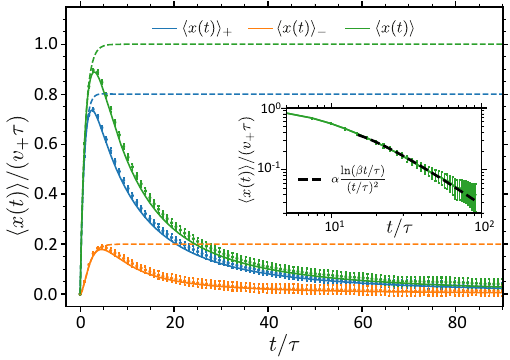}
    \caption{
    Mean position as a function of time. The main plot shows numerical data (points with error bars) for $\langle x(t)\rangle_+$, $\langle x(t)\rangle_-$ and $\langle x(t)\rangle=\langle x(t)\rangle_++\langle x(t)\rangle_-$, together with the Berezinskii solutions computed with a Pad\'e approximant (solid lines) and the classical solutions (\ref{eq:xplus_classical}) and (\ref{eq:xminus_classical}) (dashed lines). Note that not fit parameter is used. The inset shows a log-log scale of $\langle x(t)\rangle$ (points with error bars) and a fitting function $\alpha \log(\beta t/\tau)/(t/\tau)^2$ (black dashed line), with fitted parameters $\alpha=99.65$ and $\beta=0.18$.
    The initial state and parameters of the system are the same as in Fig.~\ref{fig:com_short_time}. }
    \label{fig:com_full_pade_with_class}
\end{figure}

With the short-time expansion (\ref{eq:com_quantum_short_time}) at hand, we can infer the $\langle x(t)\rangle$ using a Pad\'e approximant of the full Taylor series \cite{Baker1975}. To this aim we use that, at long time, $\langle x(t)\rangle\propto 1/t^2$ (see below). With this knowledge, we compute the mean position at any time using
\begin{equation}
    \langle x(t)\rangle=v_+\tau\left(\frac{\tau}{t}\right)^2\underset{n\to\infty}{\lim} A_n(t)
\end{equation}
where $A_n(t)$ is a diagonal Pad\'e approximant \cite{Baker1975} whose coefficients are computed from the Taylor expansion at a desired order $n$ [e.g., Eq. (\ref{eq:com_quantum_short_time}) for $n=4$]. In Fig. \ref{fig:com_full_pade_with_class} we compare the exact numerical simulations for $\langle x(t)\rangle_+$, $\langle x(t)\rangle_-$ and $\langle x(t)\rangle$ to the corresponding  Pad\'e approximants contructed from the Berezinskii technique. The plots reveal the QBE: after a few mean free times, the mean position exhibits a maximum and eventually decays to zero. For all quantities, we find a very good agreement between the simulations and the Berezinskii approach up to long times.

Let us finally come back to the long-time behavior of $\langle x(t)\rangle$. The latter is most easily visualized in the inset of Fig. \ref{fig:com_full_pade_with_class}, which shows the mean position obtained from numerical simulations of the Schr\"odinger equation in log-log scale. We find that its long-time asymptotics is well approximated by a function scaling as $\alpha \log(\beta t/\tau)/(t/\tau)^2$, which is of the same form as in spinless TRI Hamiltonians \cite{Prat2017a}. In the present case of Hamiltonian (\ref{eq:h0}), however, a direct derivation of this asymptotic limit appears to be much more involved, and is left for future work.

\section{Conclusion}

In this paper, we have extended the Berezinskii diagrammatic technique describing the dynamics of Anderson localization in one dimension to TRI-broken disordered Hamiltonians including a spin-orbit coupling term that induces an asymmetry between right and left scattering processes. As an application of the formalism, we have computed the time evolution of the mean position of a wave-packet launched in a given direction, and recovered the quantum boomerang effect discussed in \cite{Janarek2022}. As an extension of this work, it would be interesting to extract analytical long-time, asymptotic expansions for the mean position in this system, or to characterize the dynamics of other observables such as the mean square width $\langle x^2(t)\rangle$ or the full density distribution of the wave packet.

\acknowledgments

We thank Tony Prat for useful discussions on the Berezinskii technique. N.C. acknowledges financial support from the Agence Nationale de la Recherche (grant ANR-19-CE30-0028-01 CONFOCAL). J.J. acknowledges the support of French Embassy in Poland through the \emph{Bourse du Gouvernement Fran\c{c}ais} program.

\appendix

\section{Identification of scattering vertices}
\label{Appendix:vertices}

In this appendix, we briefly explain the procedure used to identify the set of phaseless scattering vertices in Fig. \ref{fig:scattering_vertices}. To this aim, let us consider the diagram in Fig.~\ref{fig:contributing_diagram}(a) once more. The diagram can be split into spatial intervals lying between consecutive scattering events $x_i$ and $x_{i+1}$. Each interval contains a specific number of lines. There are in total 4 kinds of lines: retarded lines  and advanced lines (both in two possible directions).
The numbers of lines are denoted as $g_+$ and $g_-$ (for retarded lines), and $(g_+)'$ and $(g_-)'$ (for advanced lines), with the index $\pm$ indicating their direction. For example, the interval lying between the points $x'$ and $x_3$ in the diagram from Fig.~\ref{fig:contributing_diagram}(a) has $g_+ = 2,\ g_-=1,\ (g_+)'=2,$ and $(g_-)'=1$ lines. Each scattering event induces a definite change in the number of respective lines, which we denote by $\Delta g_\pm$  and $(\Delta g_\pm)'$. These changes determine the phases associated with scattering vertices. To find these phases, we first note that each incoming and outgoing retarded propagator line at point $x$ carries a phase that depends on the direction of the line and on its type:
\begin{itemize}
    \item every incoming (outgoing) \emph{positive} line, i.e. propagating to the right, carries a $k_+x$ ($-k_+x)$ phase,
    \item every incoming (outgoing) \emph{negative} line, i.e. propagating to the left, carries a $-k_-x$ ($k_-x)$ phase.
\end{itemize} 
For advanced lines, the phases have opposite signs. 

For vertices involving only one type of lines, e.g., only retarded Green's functions, the total phase $\phi$ of a scattering vertex is then calculated from the total change of the number of lines, i.e.: 
\begin{equation}
\phi=\pm (\Delta g_+ k_+ - \Delta g_- k_-)x.
\end{equation}
Hence, the phaselessness condition of the scattering vertex in the limit of $\omega\to 0$ is that the vertex does not change the total number of incoming and outgoing lines, that is, $\Delta g_\pm = 0$. This condition is very similar to the original TRI Berezinskii method, although in our system $k_+$ and $k_-$ do not cancel each other.
In the case of the mixed-line vertices involving both $G_0^R$ and $G_0^A$, the problem is slightly different: lines from $G^R_0$ and $G^A_0$ may cancel each other. The total phase of a vertex is
\begin{equation}
    \phi = [\left(\Delta g_+ - (\Delta g_+)'\right)k_+ - \left(\Delta g_- - (\Delta g_-)'\right)k_-]x.
\end{equation}
This phase is zero only if $\Delta g_\pm - (\Delta g_\pm)' = 0$. 

\section{Differential equation for $L_m$ blocks}
\label{Appendix:Lmblock}

\begin{figure}
    \centering
    \includegraphics[width=1\linewidth]{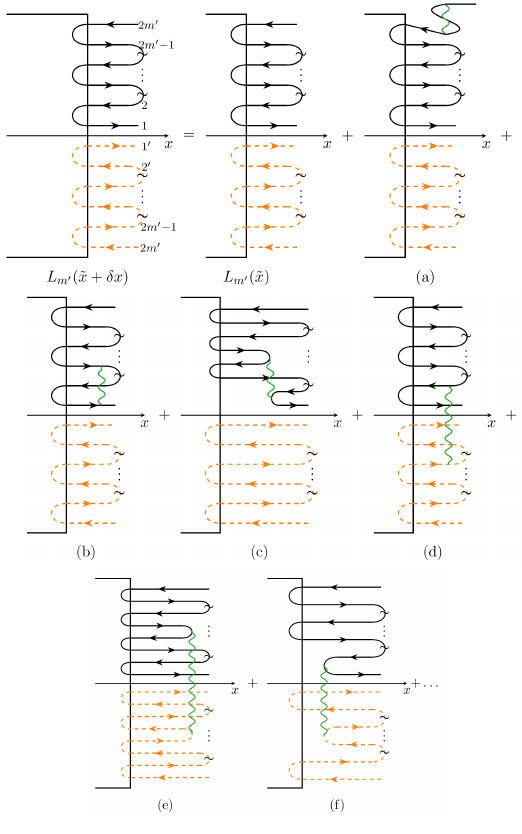}
    \caption{Schematic representation of Eq.~(\ref{spin:eq:l_m_combinatorics}). $L_{m'}(\tilde{x}+\delta x)$ can be constructed with $L_{m'}(\tilde{x})$ and all possible combinations of the scattering vertices. The figure shows one example for each scattering vertex.} \label{spin:fig:lm_construction}
\end{figure}
In this appendix, we provide details about the derivation of the differential equation (\ref{eq:l_m_full}) for the left blocks of correlation diagrams. 
To calculate $L_{m'}(\tilde{x})$, we consider how it changes with an infinitesimal change of the boundary position, say $\tilde{x}\rightarrow \tilde{x}+\delta x$ by adding all possible contributions from different scattering vertices. For this purpose, we number the lines on the boundary by assigning consecutive numbers to the outgoing and incoming lines, as presented in Fig.~\ref{spin:fig:lm_construction}. The figure also shows a schematic way of all adding new vertices to  $L_{m'}(\tilde{x})$, bearing in mind that the lines cannot create loops nor cross each other. We get the corresponding equation:
\begin{widetext}
\begin{equation}
\label{spin:eq:l_m_combinatorics}
    \begin{aligned}
        &L_{m'}(\tilde{x}+\delta x) = L_{m'}(\tilde{x}) +\frac{\eta\delta x}{2} L_{m'}(\tilde{x})\times(-2m')\left(\frac{1}{v_+}+\frac{1}{v_-}\right)^2
        \\[5pt]
        &+\eta\delta x L_{m'}(\tilde{x}) \biggl[- \frac{m'(m'-1)}{v_+^2} - \frac{m'(m'-1)}{v_-^2} -\frac{2m'^2}{v_+v_-} -\frac{2\kappa m'(m'-1)}{v_+v_-} + \frac{m'^2}{v_+^2}+\frac{m'^2}{v_-^2}+\frac{2m'^2}{v_+v_-}\biggr]\\[5pt]
        &+\frac{\eta\delta x \kappa}{v_+ v_-} \biggl[m'^2L_{m'+1}(\tilde{x})e^{i\omega \tilde{x}(\frac{1}{v_+}+\frac{1}{v_-})} + m'^2L_{m'-1}(\tilde{x})e^{-i\omega \tilde{x}(\frac{1}{v_+}+\frac{1}{v_-})}\biggr].
    \end{aligned}
\end{equation}
\end{widetext}
After taking the limit $\delta x \rightarrow 0$ and some simplifications, we finally obtain Eq. (\ref{eq:l_m_full}) of the main text.

\section{Iterative Berezinskii equations}
\label{Appendix:iterative}

In this appendix, we provide the coupled equations for for the $S_m^i$ coefficients ($i=1,2,3$) that appear in the expressions of the mean position, Eqs. (\ref{eq:com_+}) and (\ref{eq:com_-}) of the main text.

Using the same procedure as for $S_m^0$ and explained in the main text, we find the following coupled iterative equations for the $S_m^i, Q_m^i$ ($i=1,2,3$):
\begin{equation}
\label{eq:iterativeS1m}
\begin{split}
        -2\ell Q_m^1&+i\nu\left(m+\frac{v_+}{v_++v_-}\right)S_m^1-2\ell \eta\beta_m S_m^1\\
    &+m^2 S_{m-1}^1+(m+1)^2 S_{m+1}^1=0,
\end{split}
\end{equation}
\begin{equation}
\label{eq:iterativeQ1m}
\begin{split}
    \mathcal{L}_{m+1}&+i\nu\left(m+\frac{v_+}{v_++v_-}\right)Q_m^1-2\ell \eta\beta_m Q_m^1\\
    &+m^2 Q_{m-1}^1+(m+1)^2 Q_{m+1}^1=0,
\end{split}
\end{equation}
\begin{equation}
\label{eq:iterativeS2m}
\begin{split}
    2\ell Q_m^2&+i\nu\left(m+\frac{v_-}{v_++v_-}\right)S_m^2-2\ell \eta\beta_m S_m^2\\
    &+m^2 S_{m-1}^2+(m+1)^2 S_{m+1}^2=0,
\end{split}
\end{equation}
\begin{equation}
\label{eq:iterativeQ2m}
\begin{split}
        \mathcal{L}_{m+1}&+i\nu\left(m+\frac{v_-}{v_++v_-}\right)Q_m^2-2\ell \eta\beta_m Q_m^2\\
    &+m^2 Q_{m-1}^2+(m+1)^2 Q_{m+1}^2=0,
\end{split}
\end{equation}
\begin{equation}
\label{eq:iterativeS3m}
\begin{split}
    -2\ell Q_m^3&+i\nu\left(m+\frac{v_+}{v_++v_-}\right)S_m^3-2\ell \eta\beta_m S_m^3\\
    &+m^2 S_{m-1}^3+(m+1)^2 S_{m+1}^3=0,
\end{split}
\end{equation}
\begin{equation}
\label{eq:iterativeQ3m}
\begin{split}
    \mathcal{L}_{m+1}&+i\nu\left(m+\frac{v_+}{v_++v_-}\right)Q_m^3-2\ell \eta\beta_m Q_m^2\\
    &+m^2 Q_{m-1}^3+(m+1)^2 Q_{m+1}^3=0.
\end{split}
\end{equation}

\section{Partial components $\langle x(t)\rangle_\pm$}
\label{Appendix:short-time}

We finally provide the short-time expansions for $\langle x(t)\rangle_+$ and $\langle x(t)\rangle_-$, and the exact expressions (valid at any time) of their classical counterparts $\langle x(t)\rangle^{\text{class.}}_+$ and $\langle x(t)\rangle^{\text{class.}}_-$:
\begin{align}
\label{eq:xplus_quantum}
        &\frac{\langle x(t)\rangle_+}{v_+\tau} =  \left[\frac{t}{\tau} - \frac{1}{1+\Delta}\frac{t^2}{\tau^2} +  \frac{3-\Delta}{6(1+\Delta)}\frac{t^3}{\tau^3} \right.  \\
        &\left. - \frac{\Delta(\Delta+1)(\Delta(\Delta^2+\Delta-3)-7)-2}{12(\Delta+1)^5}\frac{t^4}{\tau^4}\right] + \mathcal{O}(t^5),\nonumber
\end{align}
\begin{align}
\label{eq:xminus_quantum}
        &\frac{\langle x(t)\rangle_-}{v_+\tau} =  \left[\frac{1-\Delta}{2(1+\Delta)}\frac{t^2}{\tau^2} - \frac{1-\Delta}{3(1+\Delta)}\frac{t^3}{\tau^3} +  \right.  \\
        &\left. - \frac{\Delta(9-\Delta(\Delta(3\Delta(\Delta+3)+8)-8))+3}{24(\Delta+1)^5}\frac{t^4}{\tau^4}\right] + \mathcal{O}(t^5),\nonumber
\end{align}
\begin{align}
\label{eq:xplus_classical}
    \frac{\langle x(t)\rangle_{+}^{\text{class.}}}{\tau v_+} \!=&  \biggl[ \frac{2\Delta}{1 + \Delta}\bigl(1 - e^{-t/\tau}\bigr) + \frac{1 - \Delta}{1 + \Delta} \frac{t}{\tau}e^{-t/\tau} \biggr],
\end{align}
\begin{align}
\label{eq:xminus_classical}
    \frac{\langle x(t)\rangle_{-}^{\text{class.}}}{\tau v_+}\! =&  \frac{1- \Delta}{1+\Delta}\biggl(1 - e^{-t/\tau} - \frac{t}{\tau}e^{-t/\tau}\biggr).
\end{align}
\newline

\bibliography{biblio}

\end{document}